\documentclass[10pt,twocolumn,twoside]{IEEEtran}
\usepackage{times}
\usepackage{cite}
\usepackage{color}
\usepackage{epsf}
\usepackage{epsfig}
\usepackage{graphicx}
\usepackage{graphics}
\usepackage{amsmath}
\usepackage{amssymb}
\usepackage{amsthm}
\usepackage{amsxtra}
\usepackage{algorithm}
\usepackage{algorithmic}
\usepackage{enumerate}
\usepackage{multirow}
\usepackage{enumerate}
\usepackage{amssymb}
\usepackage{lipsum}
\usepackage{setspace}
\usepackage{multirow}

\usepackage{mdwmath}
\usepackage{mdwtab}
\usepackage{eqparbox}
\usepackage{wrapfig}
\usepackage{rotating} 
\usepackage{graphicx} 

\usepackage{url}

\newcommand{\Rmnum}[1]{\expandafter\@slowromancap\romannumeral #1@}

\makeatother
\begin{document}
\title{Demand Response Management For Power Throttling Air Conditioning Loads In Residential Smart Grids}


\author{Yawar Ismail Khalid, Naveed Ul Hassan,~\IEEEmembership{Member,~IEEE,}  ~Chau~Yuen,~\IEEEmembership{Senior Member,~IEEE} \\ and Shisheng~Huang,~\IEEEmembership{Member,~IEEE}
\thanks{Y. I. Khalid and N. U. Hassan are with the Electrical Engineering Department at Lahore University of Management Sciences (LUMS), Lahore, Pakistan 54792. (Email: yawar89@gmail.com, naveed.hassan@yahoo.com).}
\thanks{C. Yuen and S. Huang are with the Engineering Product Development at the Singapore University of Technology and Design (SUTD), Dover Drive, Singapore 138682. (Email: \{yuenchau, shisheng\_huang\}@sutd.edu.sg).}
\thanks{This research is partly supported by Lahore University of Management Sciences (LUMS) Research Startup Grant and Energy Innovation Research Program (EIRP) Singapore NRF2012EWT-EIRP002-045.}
}
\IEEEoverridecommandlockouts
\maketitle


\begin{abstract}
In this paper we develop an algorithm for peak load reduction to reduce the impact of increased air conditioner usage in a residential smart grid community. We develop Demand Response Management (DRM) plans that clearly spell out the maximum duration as well as maximum severity of inconvenience. We model the air conditioner as a power throttling device and for any given DRM plan we study the impact of increasing the number of power states on the resulting peak load reduction. Through simulations, we find out that adding just one additional state to the basic ON/OFF model, which can throttle power to 50\% of the rated air conditioner power, can result in significant amount of peak reduction. However, the peak load that can be reduced is diminishing with the increase in number of states. Furthermore, we also observe the impact of inconvenience duration and inconvenience severity in terms of peak load reduction. These observations can serve as useful guidelines for developing appropriate DRM plans.

\end{abstract}

\section{Introduction}
The traditional electrical grid relied solely on adjusting it's generation in response to consumer load variations. Recent influx of smart grid technologies has allowed for a bi-directional communication and power flow, making it possible to achieve the supply demand balance by modifying the demand side of the equation. Demand Response Management (DRM), thus refers to regulating and shaping the demand to match supply \cite{DRM_1, DRM_2, DRM_3, neww1, neww2, neww3, neww4, neww5}. This is crucial because in peak demand hours grids call on peaking power plants, which are very costly to operate, to supply the additional electrical power. The overall increase in electricity price is mainly due to the increasing cost of peak energy \cite{elec_price1, elec_price3}. DRM is also important in order to avoid blackouts during instances of insufficient generation and to account for the variable nature of generation from renewable energy sources like wind and solar. 

Increased peak electricity demand has become a major problem for today's grid since designing a system with a capacity to meet infrequent and short periods of high electricity demand requires a disproportionate share of power generation and network investment. A number of studies have attributed this peak demand increase to the increased usage of air conditioners especially in summer months. In \cite{koomey} the authors claim that peak demand in the warmer regions of the US ``is driven mainly by air conditioning loads on the hottest summer afternoon". In \cite{tokyo} it is reported that the peak demand in Japan occurs due to increased usage of air conditioners in the summers. \cite{ees} states that the peak demand in Australia and New Zealand always coincided with the high outdoor temperature days, and that this was due to the increased usage and penetration of air conditioners in the region. \cite{hongkong} reports that the average annual electricity consumption pattern by end user appliances has a 46\% contribution from air conditioners in Hong Kong (a figure that increases to 59.1\% in summers), clearly pointing out air conditioners as the main cause of increased grid peak load.

A number of efforts have been made to reduce peak load due to increased air conditioning usage in summer months. In \cite{survey}, the authors outline a number of case studies where DSM has been used to reduce peak electricity consumption. The Australian and New Zealand Governments jointly formed AS/NZS 4755 standard \cite{aus_stand} which mandates physical/electrical interface as well as mandatory and optional DRM modes for air conditioners being manufactured and sold in Australia and New Zealand. DRM modes permit the air conditioners to operate at variable power levels. Variable frequency and variable speed drives can be used to throttle power and switch the air conditioner operation between various DRM modes. 

  
Typically air conditioning load can be controlled either by adjusting its thermostat or by adjusting the compressor power also called DRM modes. In the first method, a smart thermostat is installed in customer's premises to automatically control the air conditioner thermostat. Google Nest is one such example, which can adjust the thermostat set point and the air conditioner then consumes power according to the setting \cite{googlenest}. In this method, the lower the thermostat set point, the higher is the amount of power consumed. In the second method, DRM mode is controlled to alter the state of the air conditioner. This method requires an interface with the device to receive the control signal representing the state of the device and then throttle the power accordingly. In a basic two state model, the air conditioner can only be turned OFF/ON where, OFF state is the zero power state while ON state is the rated power state. Companies such as Idaho Power have introduced AC Cool Credit program, which allows them to switch OFF/ON the air conditioners of their customers for a certain amount of time according to an agreement and report significant peak load reduction \cite{idaho}. When the air conditioner is turned OFF, the customers can experience unbounded variations away from their desired thermostat set point.

In this paper, we consider power throttling type of control for the air conditioners. Instead of just considering a basic 2-state model, we develop a generic $K$-state model as mandated by \cite{aus_stand}. For example, in a 3-state model, air conditioner can be turned OFF, ON or operated at 50\% of the rated power. We study the problem of peak load reduction by controlling the air conditioners in a residential smart grid community, while considering DRM plans which are easily comprehensible. Our DRM plans thus clearly specify the ``maximum inconvenience duration" (maximum time duration in which air conditioner is denied demanded operation at full rated power in a particular day) as well as ``maximum inconvenience severity" (maximum temperature variation from the thermostat set point). We develop an algorithm that tells us the effectiveness of DRM plans and throttlable states in reducing the peak load on the grid. We simulate various DRM plans and the amount of peak reduction each plan can offer with different number of throttlable states. Simulation results show that with a basic 2-state ON/OFF model, peak load reduction of up to 16.5\% can be achieved with a DRM plan with maximum inconvenience duration of 1 hour and maximum inconvenience severity of 3 $^oF$ from the thermostat set point of 65 $^oF$. The amount of peak load reduction increases to 21\% when a third state capable of operation at 50\% rated power is added. However, increasing the number of states to 5 yields marginal returns over a 3-state model resulting in 22\% peak load reduction. We also study the impact of maximum inconvenience duration and maximum inconvenience severity parameters on peak load reduction. The results in this paper can provide useful guidelines for developing appropriate DRM plans, incentives and financial rewards for the smart grid users.

 
The rest of the paper is organized as follows. The system model, load model, power consumption and thermal model of air conditioner and description of DRM plans is given in Section II. The optimization problem and DRM algorithms are given in Section III. Simulation results are presented in Section IV while the paper is concluded in Section V.

\section{System Model, Load Model, Power Consumption and Thermal Load Model of Air conditioner and DRM Plans}

\subsection{System Model} 
We consider a smart grid community comprising of $J$ homes (also called users, consumers, customers) where each home is equipped with an air conditioner. The air conditioner is assumed to be capable of power throttling and can operate in $K$ different states. We assume a uni-directional power flow from the grid to the consumers. We assume a physical/electrical interface (also called demand response enabling device) \cite{aus_stand} connected to the air conditioner capable of bi-directional information exchange with the controller. This interface is also assumed to be capable of operating the air conditioner in the desired state according to the DRM signal received from the grid. 
In our system model, there is a grid controller that is assumed to be connected to all the homes in the residential community. In an open setup, the grid controller is assumed to be communicating directly with the air conditioners. In this setup it is also assumed that the grid controller knows the power consumption profile as well as usage pattern of each air conditioner in each home. In a more private setup, a home controller can be installed in each home which can then act as an additional interface between the grid controller and the air conditioner. In this setup, home controller communicates with the air conditioner according to the commands received from the grid controller. The power consumption profile and usage pattern of the air conditioner is known only to the home controller. It should be noted that the presence of a home controller in the system model only ensures privacy of individual household data. Moreover, since we are considering only one flexible load in this paper, therefore, the functionality of a home controller and the physical interface on the air conditioner might also be thought of as a single interface. 

\subsection{Load Model}
We consider the power consumption and usage patterns of a smart grid community over a period of 24 hours at a granularity of 5 minutes (hence yielding a total of $T$ = 288 time slots in a day). For the purposes of this paper we consider all non-air conditioning loads to be essential loads, which have fixed scheduling needs and fixed power consumption. The grid cannot control these loads in any possible way and is required to provide them with the necessary power at the exact time of their operation. Examples of such loads include refrigerators, televisions, computers, lights etc. 

\subsection{Power Consumption Model of Air conditioner}
This model is required in order to compute the power consumption of air conditioner. Let, $P^{AC}_{j}$ denote the rated power of air conditioner of user $j$. At any time instant $t$ the air conditioner is assumed to be operating in one of the possible states $k\in[1,K]$. In subsequent discussion $k=1$ will always correspond to the state in which the air conditioner is OFF, whereas $k=K$ will always correspond to the state in which the air conditioner is being operated at the rated power. The power $p^{AC}_{j,k}(t)$ consumed by the air conditioner of user $j$ in state $k$ is then given by,
\begin{equation}
p^{AC}_{j,k}(t)=\frac{k-1}{K-1}\times P^{AC}_j, \quad k = 1,...,K                
\label{power_levels}
\end{equation}
For example, an air conditioner that can operate in 5-states has the capability to throttle power at 0\%, 25\%, 50\%, 75\% and 100\% of the rated power. 

\subsection{Thermal Load Model of Air conditioner}
The room temperature variations resulting from throttling air conditioner power can be estimated using a thermal load model of air conditioner. 
Let $\alpha_j$ and $\beta_j$ denote respectively the start time and end time of the interval in which user $j$ has demanded air conditioner. Let $W^{AC}_{j}(t)$ denote a binary variable indicating the demand status of user $j$'s air conditioner in time slot $t$,
\begin{center}
\begin{equation}
W^{AC}_{j}(t)=\Bigg\{\begin{array}{c}
                1,\quad \forall t\in [\alpha_j, \beta_j] \\
                0,\quad \quad \:  \text{otherwise}
                \end{array}
                \label{myc2}
\end{equation}
\end{center}
Similarly, let $C^{AC}_{j,k}(t)$ denote a binary variable indicating whether air conditioner of user $j$ is operating in state $k$ or not at time $t$,
\begin{center}
\begin{equation}
C^{AC}_{j,k}(t)=\Bigg\{\begin{array}{c}
                0,\quad \text{if AC is not operating in state k}\\
                1,\quad \text{if AC is operating in state k}
                \end{array}
                \label{myc3}
\end{equation}
\end{center}
It should also be noted here that the air conditioner can be operated in only one state at a particular time. In other words, it is impossible for an air conditioner to be simultaneously in two or more states. Let, $\theta^{AC}_{j,k}(t)$ denote the room temperature of user $j$ at time $t$ when the air conditioner is operating in state $k$. The resulting room temperature at the start of the next time slot $t+1$ can then be measured using a model similar to the one developed in \cite{g14},
\begin{equation} 
\theta^{AC}_{j,k}(t+1) = \theta^{AC}_{j,k}(t) + \Delta{t} \frac{G_j(t)}{\Delta{c}} + \Delta{t} \frac{\hat{Z}^{AC}_{j,k}}{\Delta{c}}{W}^{AC}_{j}(t)C^{AC}_{j,k}(t)
\label{Ti+1:hvac}
\end{equation}
In this equation, $G_j(t)$ is the heat gain rate of the house of user $j$ which depends on heat gain coefficients of the walls, windows, roof, solar radiation, people and air change rate of the AC etc., and is independent of the state $k$ in which the air conditioner is operating, $\Delta{t}$ is the granularity of a time slot while $\Delta{c}$ is the energy required for a unit degree rise in room temperature. The parameter $\hat{Z}^{AC}_{j,k}$ in this model is the cooling capacity of the air conditioner when it is operating in state $k$. The cooling capacity is given in BTU/hr and is a function of the state $k$ in which the air conditioner is operating. Generally, the cooling capacity is directly proportional to the power at which the air conditioner is being operated, 
\begin{equation}
\hat{Z}^{AC}_{j,k} = \textit{EER} \times p^{AC}_{j,k} 
\end{equation}
where $p^{AC}_{j,k}$ (power consumption of air conditioner when it's operating in state $k$) is given in kW and $EER$ is the energy efficiency ratio of the air conditioner. The US national appliance standards dictate all ACs to have a minimum value of $EER \geq 8.0$ \cite{energygov}.

\subsection{DRM Plan}
We propose to offer DRM plans to residential customers, which are not only easy to comprehend, but also guarantee a bounded maximum inconvenience. For an air conditioning load, inconvenience has two dimensions i.e. duration and severity. We define inconvenience duration as the total time in which the air conditioner is demanded by the user but is denied operation at the full rated power. Similarly we define inconvenience severity as the temperature deviation from the thermostat set point. The maximum inconvenience in our DRM plan is therefore specified in terms of maximum inconvenience duration and maximum inconvenience severity. The grid will then ensure that the inconvenience experienced by any user does not exceed the specified values in both the dimensions. 
Furthermore, in order to ensure fairness, we also make the desired thermostat set point part of the DRM plan. Thus any DRM plan consists of three values, 
\begin{enumerate}
	\item $\hat{\theta}^{AC}$: denotes the thermostat set point. 
	\item $\Delta{\theta^{AC}}$: denotes the maximum temperature deviation from the thermostat set point.
	\item $\tilde{t}^{AC}_{max}$: denotes the maximum time duration during the demanded interval in which thermostat set point temperature is not provided to the user.
\end{enumerate}
Customers can therefore easily comprehend and subscribe to a DRM plan that suites their needs. For the grid operator, it then becomes essential to determine mutually beneficial DRM plans that not only serve the grid (in terms of peak load reduction) but also benefit the customers (in terms of incentives offered by the grid for facing inconvenience\footnote{Designing incentives is not our focus in this paper}).

\section{Problem Formulation and Algorithm Design}
In this section we formulate an optimization problem in order to determine the amount of peak reduction that can be achieved using any given DRM plan. Let $\bar{E}$ denote the peak load on the grid (without DRM) due to the aggregate load demand of the residential community. For a given DRM plan we have the following optimization problem,
\begin{equation}
\begin{aligned}
& \underset{C^{AC}_{j,k} }{\text{minimize}}
& & \bar{E} \\
\end{aligned}
\end{equation}
subject to the following constraints,
\begin{center}
\begin{equation}
W^{AC}_{j}(t) \left( \theta^{AC}_{j,k}(t) - \hat{\theta}^{AC}\right) \leq \Delta{\theta^{AC}}, \: \forall t, \forall j, \forall k
\label{third}
\end{equation}
\end{center}
\begin{center}
\begin{equation}
\sum_{t=1}^{T} W^{AC}_{j}(t)- \sum_{t=1}^{T} W^{AC}_{j}(t)C^{AC}_{j,k=K}(t) \leq \tilde{t}^{AC}_{max}, \forall j
\label {fourth}
\end{equation}
\end{center}
\begin{equation}
\sum_{k=1}^{K} C^{AC}_{j,k}(t)=1, \forall t, \forall j
\label{fifth}
\end{equation}
The objective is to minimize peak load on the grid by determining the optimization variables $C^{AC}_{j,k}(t)$ i.e. the throttling state $k$ in which the air conditioner of each user $j$ should be operated. Constraint \eqref{third} bounds inconvenience severity and requires that the deviation of actual room temperature in the demanded operation interval of any user does not exceed the thermostat set point by more than $\Delta{\theta^{AC}}$. Constraint \eqref{fourth} bounds the inconvenience duration and requires that the number of time slots in which user $j$ is denied service at the full rated power is less than or equal to $\tilde{t}^{AC}_{max}$. Constraint \eqref{fifth} is required to ensure that the air conditioner cannot operate simultaneously in two or more states i.e. air conditioner can be in any one state $k$ in any given time slot $t$. In the rest of the paper we will refer to constraint \eqref{third} as ``inconvenience severity constraint", while constraint \eqref{fourth} as ``inconvenience duration constraint".

In the subsequent discussion we will assume a home controller in each home (however, the algorithm and the solution will remain unchanged if we combine the functionality of a home controller and the physical interface with the air conditioner or even if we consider an open set up). This problem is NP hard due to the fact that the order in which different homes are considered affect the solution. Therefore finding an optimal solution is not practical and has exponential complexity. Fortunately, optimal solution is also not required for this problem since using any random order can give a fairly good idea about the effectiveness of DRM plan as well as the effect of throttlable states on peak reduction.

We develop an offline algorithm (Algorithm \ref{alg:seq}) for peak load reduction for any given DRM plan. This algorithm assumes the prior knowledge of the aggregated load profile of the residential community. The grid controller can obtain this information either from past power consumption patterns of the community during the same time period or it can use some prediction models to accurately forecast both short term and long term load demand on the grid \cite{forecast1,forecast4}. 
Below we explain our algorithm.\\
\textbf{Sequential Algorithm (Algorithm \ref{alg:seq})}: The grid controller broadcasts the DRM plan and then sequentially asks each home controller to contribute towards peak load reduction by experiencing the inconvenience according to the DRM plan. The grid controller sends the aggregated load profile of the community to the home controller of user 1 (in any random order). At the home controller Algorithm \ref{alg:home} is implemented which determines the appropriate throttling states of air conditioner in its demanded operation time such that the inconvenience constraints laid out in DRM plan are not violated. The modified aggregated load profile is reported back to the grid controller by the home controller of user 1. The grid controller sends the new load profile to the home controller of user 2. The grid controller continues this process sequentially for all the users in the residential community and obtains the final aggregated load profile with reduced peak load. \\
\begin{algorithm}[htb]
\caption{Sequential Algorithm}
\label{alg:seq}
\begin{algorithmic}[1]
\STATE Grid broadcasts the DRM plan to all the home controllers in the residential community. 
\STATE For $j=1:J$
\STATE Grid controller sends the aggregated load profile of the community to user $j$.
\STATE User $j$ executes the home control algorithm \ref{alg:home}.
\STATE User $j$ reports the new aggregated load profile to the grid controller.
\STATE End for
\STATE Grid has the final aggregated load profile indicating the total peak load reduction.
\end{algorithmic}
\end{algorithm}
\textbf{Home Control Algorithm (Algorithm \ref{alg:home})}:The home controller of user $j$ receives the aggregated load profile denoted by $E(t), \: \forall t$ from the grid controller and executes home control algorithm (Algorithm \ref{alg:home}). Let $\tilde{t}_j$ denote a counter initialized to zero. Similarly we initialize all the state indicator variables $C^{AC}_{j,k}(t)$ of user $j$ equal to zero. Let $t_{dur}=\frac{\tilde{t}^{AC}_{max}}{\Delta t}$ denote the maximum number of time slots during which the operation of air conditioner can be denied at the rated power. Every time user is denied operation at its rated power during its demanded operation interval $t\in [\alpha_j, \beta_j]$ this counter is incremented by $\Delta t$ (i.e. the duration of one time interval). If $\tilde{t}_j < \tilde{t}^{AC}_{max}$, since the inconvenience duration as laid out in the DRM plan is not yet exhausted, the home controller determines the time index in the interval $t\in [\alpha_j, \beta_j]$ where the local peak occurs i.e. 
\[t_j^p=\max_{t\in [\alpha_j, \beta_j]} E(t) \]
In this time slot $t_j^p$, the home controller generates an appropriate DRM control signal for the physical interface in order to operate the air conditioner in a state that leads to maximum peak reduction without letting the room temperature deviate more that $\Delta{\theta^{AC}}$ from the thermostat set point. The home controller uses the thermal model of the air conditioner \eqref{Ti+1:hvac} to determine the optimal power state denoted by $k^*$. The home controller starts by considering switching OFF the air conditioner (since it can result in maximum power reduction). If the resulting room temperature at the start of next time interval is within the allowable limit, then this is the optimal state and the loop is terminated. However, if the room temperature exceeds the thermostat set point by $\Delta{\theta^{AC}}$, the home controller considers operating the air conditioner in the next lowest power state. This process continues until the optimal state of the air conditioner is determined. The power consumption variable and total load value are updated at the end of this operation. The algorithm then starts again by finding new local peak and steps are repeated. Finally the new aggregated load profile is communicated by the home controller of user $j$ to the grid controller.
This algorithm will either: 
\begin{itemize}
	\item \textbf{achieve the inconvenience duration constraint with strict equality}: in which case the user will experience the maximum duration of inconvenience as spelled out in the DRM plan without violating the inconvenience severity constraint.
	\item \textbf{achieve the inconvenience duration constraint with strict inequality}: in which case the inconvenience duration is strictly less than the maximum inconvenience duration. This situation can arise if under some given environmental conditions, operation of air conditioner in any other state $k\neq K$ violates the inconvenience severity constraint. In such a situation, air conditioner operation at only the full rated power, can restrict the room temperature deviation to a value less than or equal to $\Delta{\theta^{AC}}$. This case can provide useful guidelines to design appropriate DRM plans. For example, if a DRM plan fails to exhaust the inconvenience duration of several users then it indicates that inconvenience duration of the plan can be reduced without having a significant impact on the peak load reduction.
\end{itemize}

\begin{algorithm}[htb]
\caption{Home Control Algorithm}
\label{alg:home}
Initialize: $\tilde{t}_j=0$, \\
$C^{AC}_{j,k}(t)=0, \forall t,k$
\begin{algorithmic}[1]
\FOR{$i=1:t_{dur}$}
\STATE Increment: $\tilde{t}_j=\tilde{t}_j+\Delta t$
\STATE Determine time index: $t_j^p=\max_{t\in [\alpha_j, \beta_j]} E(t)$
\FOR{$k=1:K$}
\STATE Set: $C^{AC}_{j,k}(t_j^p)=1$
\STATE Using \eqref{Ti+1:hvac}, determine $\theta^{AC}_{j,k}(t_j^p+1)$ 
\IF{$\left(\theta^{AC}_{j,k}(t_j^p+1)-\hat{\theta}^{AC} \right) \leq \Delta{\theta^{AC}}$}
\STATE The optimal state is: $k^*=k$
\STATE break for loop
\ELSE 
\STATE Set: $C^{AC}_{j,k}(t_j^p)=0$
\ENDIF
\ENDFOR
\STATE Determine the power: $p^{AC}_{j,k^*}(t_j^p)$ using \eqref{power_levels}
\STATE Update: $E(t_j^p)=E(t_j^p)-\left(P^{AC}_{j}-p^{AC}_{j,k^*}(t_j^p)\right)$
\ENDFOR
\STATE Report the new aggregated profile to the grid controller.
\end{algorithmic}
\end{algorithm}

\section{Case Study and Simulation Results}
\begin{figure}[htb]
\centering
\includegraphics[width=0.45\textwidth,height=.18\textheight]{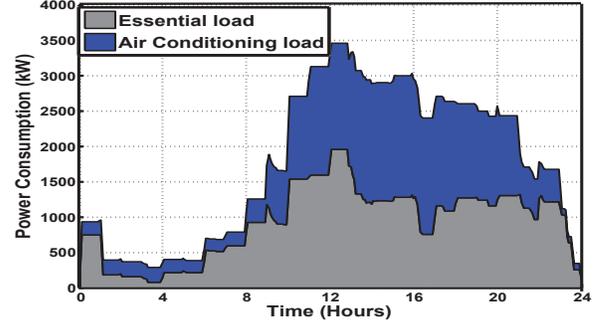}
\caption{Aggregated load profile of a residential community}
\label{fig:fig001}
\end{figure}

In this section we analyze the performance of our designed algorithm through a case study. We consider a smart grid community comprising of 1000 homes. The air conditioning and other essential loads of these homes are generated using realistic appliance usage and power consumption patterns as given in \cite{energy1}-\cite{energy_pattern}. The average daily household energy consumption is assumed to be about 41 kWh, which corresponds to typical household energy consumption in many US states like Louisiana, Tennessee, Alabama etc. The air conditioner in each home is assumed to be operated for a total of four consecutive hours. The operation time of the air conditioners, however vary from user to user, with a high number of users demanding operation between 1:00 P.M and 6:00 P.M. In this case study, we assume that the rated capacity of air conditioner in each home is approximately 4.2 tons with an EER value of 10, giving a rated power consumption of 5 kW. Figure \ref{fig:fig001} shows the aggregated load profile of the 1000 homes on which DRM algorithms are implemented. It clearly shows the contribution of air conditioners towards peak load increase during the summer months. For our generated load profile, peak load value is 3458 kW. The room temperature is calculated according to \eqref{Ti+1:hvac}. 

 \begin{table}[htb]
 \caption{DRM Plans and resulting percentage peak reduction for $K$=2, $K$=3 and $K$=5 power throttling states}
 \centering
 \begin{tabular} {|c|c| c| c| c|c|  }
 \cline{1-6}
  & & & \multicolumn{3}{ c| }{Percentage peak reduction (\%)} \\ \cline{4-6}
 \multirow{1}{*}{$\hat{\theta}^{AC}$}&\multirow{1}{*}{$\Delta{\theta}^{AC}$}&\multirow{1}{*}{$\tilde{t}^{AC}_{max}$} &\multirow{1}{*}{$K$=2}  &  \multirow{1}{*}{$K$=3}&  \multirow{1}{*}{$K$=5}\\ 
 \cline{1-6}
\multirow{1}{*}{}&\multirow{1}{*}{}& \multirow{1}{*}{1 hour}& \multirow{1}{*}{16.5\%} &\multirow{1}{*}{21.1\%}  &  \multirow{1}{*}{21.9\%}\\
 \cline{3-6}
 \multirow{1}{*}{65$^oF$}&\multirow{1}{*}{3$^oF$}& \multirow{1}{*}{1.5 hours}& \multirow{1}{*}{16.9\%} &\multirow{1}{*}{21.4\%}  &  \multirow{1}{*}{22.1\%}\\
  \cline{3-6}
 \multirow{1}{*}{}  &\multirow{1}{*}{}& \multirow{1}{*}{2 hours}& \multirow{1}{*}{17.5\%} &\multirow{1}{*}{21.8\%}  &  \multirow{1}{*}{22.5\%}\\
  \cline{1-6}
 \multirow{1}{*}{}& \multirow{1}{*}{}& \multirow{1}{*}{1 hour}& \multirow{1}{*}{20.0\%} &\multirow{1}{*}{23.4\%}  &  \multirow{1}{*}{23.8\%}\\
  \cline{3-6}
  \multirow{1}{*}{65$^oF$}&\multirow{1}{*}{5$^oF$}& \multirow{1}{*}{1.5 hours}& \multirow{1}{*}{20.2\%} &\multirow{1}{*}{23.9\%}  &  \multirow{1}{*}{24.3\%}\\
   \cline{3-6}
  \multirow{1}{*}{}&\multirow{1}{*}{}& \multirow{1}{*}{2 hours}& \multirow{1}{*}{20.8\%} &\multirow{1}{*}{25.6\%}  &  \multirow{1}{*}{26.1\%}\\
   \cline{1-6}
    
   \end{tabular}
 \label{results}
 \end{table}

We simulate six different DRM plans which are summarized in Table \ref{results}. Each plan explicitly specifies the thermostat set point, maximum inconvenience duration as well as maximum inconvenience severity. In all the simulated DRM plans we assume same value of thermostat set point i.e. $\hat{\theta}^{AC}$ = 65$^oF$. In one set of DRM plans the maximum inconvenience severity is kept at 3$^oF$ while the maximum inconvenience duration is varied. Similarly in the second set of DRM plans, the maximum inconvenience severity is kept at 5$^oF$ while the maximum inconvenience duration is varied. We simulate all these DRM plans for different set of throttling states. We simulate 2-state (OFF, rated power), 3-state (OFF, 0.5 $\times$ rated power, rated power) and 5-state (OFF, 0.25 $\times$ rated power, 0.5 $\times$ rated power, 0.75 $\times$ rated power, rated power) models. The percentage peak reduction each DRM plan can achieve for different power states is also given in Table \ref{results}. We have the following observations:\\
\textbf{Observation 1:} For any given DRM plan, when the throttling power states of air conditioner are increased the amount of peak load reduction is also increased. For example, when $\Delta{\theta}^{AC}=$3$^oF$ and $\tilde{t}^{AC}_{max}$=1 hour, $K$=2 achieves 16.5\% peak reduction compared to 21.1\% for $K$=3 and $21.9\%$ for $K$=5. \\
\textbf{Observation 2:} Increasing the number of throttling states from 2 to 3 always lead to significant gains in terms of peak load reduction. However, the return is marginal if the number of states are further increased. Thus adding only one more throttling state in the basic 2-state model can lead to significant reduction in peak load. \\
\textbf{Observation 3:} For a given value of $\tilde{t}^{AC}_{max}$, increasing the maximum inconvenience severity leads to more peak reduction. For example, when $\tilde{t}^{AC}_{max}$=1 hour and $K=2$, setting $\Delta{\theta}^{AC}=$5$^oF$ results in 20\% peak reduction compared to 16.5\% peak reduction when $\Delta{\theta}^{AC}=$3$^oF$. \\
\textbf{Observation 4:} For a given value of $\Delta{\theta}^{AC}$, increasing the maximum inconvenience duration results in more peak load reduction for any number of throttling states of the air conditioner. However, we can also observe that increasing the maximum inconvenience duration only leads to a slight or even no increase in peak load reduction. This is due to the fact that in the simulated DRM plans the values of $\tilde{t}^{AC}_{max}$ are quite high and it is not possible to exhaust the maximum inconvenience duration without violating the maximum inconvenience severity constraint. Thus maximum inconvenience duration constraint in these plans is achieved with strict inequality. \\
\textbf{Observation 5:} These results also suggest the existence of saturation limits on $\Delta{\theta}^{AC}$ and $\tilde{t}^{AC}_{max}$ variables. Setting the inconvenience parameters above the saturation limits, do not offer any significant peak reduction.

\begin{figure}[htb]
\centering
\includegraphics[width=.5\textwidth,height=.18\textheight]{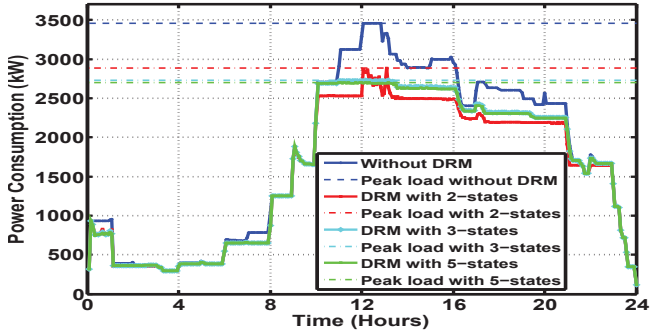}
\caption{Aggregated load profile with and without DRM. $\Delta{\theta}^{AC}$ = 3$^oF$ and $\tilde{t}^{AC}_{max}$ = 1 hour for $K$=2, $K$=3 and $K$=5 power throttling states} 
\label{fig:fig002}
\end{figure}

In Figure \ref{fig:fig002} we plot the aggregated load profiles that are obtained for DRM plan with $\Delta{\theta}^{AC}$ = 3$^oF$ and $\tilde{t}^{AC}_{max}$ = 1 hour for $K$=2, $K$=3 and $K$=5 power throttling states. Again this figure shows that increment of just one more power throttling state can result in significant peak load reduction compared to the basic ON/OFF control. Furthermore, having a more fine control only results in marginal gains.  

\section{Conclusion}
In this paper we developed a DRM algorithm for peak load reduction due to increased air conditioner usage in summer months. We proposed DRM plans, which clearly describe the deviation of room temperature from the thermostat set point as well as the maximum duration of such deviations in the demanded operation interval of the air conditioner. We modeled the air conditioner as a power throttling device. For a given DRM plan, we studied the impact of power throttling on peak load reduction. A basic 2-state ON/OFF model leads to significant reduction in peak load. Moreover, adding just one additional state can result in significant further peak load reduction. However, having a much finer control with more than 3-states results in marginal gains. We also study the impact of maximum inconvenience duration and maximum inconvenience severity parameters on peak reduction. In future, it would be interesting to consider the combined impact of load shifting and power throttling, as well as thermal modeling inaccuracies on peak reduction. Designing mutually beneficial incentives to encourage user participation through DRM plans is also an interesting future work.

\end{document}